\documentclass{ws-procs9x6}
\usepackage{amssymb}
\usepackage{amsfonts}
\ifx\cal\undefined
    \def\cal #1{{\mathcal #1}}
\fi

\begin{document}

\title{NUCLEAR MANY-BODY PHYSICS WHERE STRUCTURE AND REACTIONS MEET\footnote{\uppercase{T}he collaboration with \uppercase{V}.
\uppercase{Z}elevinsky is highly appreciated. \uppercase{T}he
work was supported by the \uppercase{U}. \uppercase{S}. \uppercase{D}epartment of \uppercase{E}nergy, grant \uppercase{DE-FG02-92ER40750}.}}

\author{Naureen Ahsan and Alexander Volya}

\address{Department of Physics,
Florida State University, Tallahassee, FL 32306, USA}

\maketitle

\abstracts{
The path from understanding a simple reaction problem of scattering
or tunneling to contemplating the quantum nuclear many-body system,
where structure and continuum of reaction-states meet, overlap and
coexist, is a complex and nontrivial one. In this presentation we
discuss some of the intriguing aspects of this route. 
}

\section{Introduction}

Structure and reactions have traditionally been separate subjects
of nuclear physics; however, recently the need for unification of our
approach to open many-body systems has become apparent. The advances
in experimental techniques led to observation of exotic nuclei which
exist only due to complex interplay between many-body structural effects
and reaction dynamics. Open mesoscopic systems like microwave cavities,
quantum dots, nuclei, and even hadrons, play an increasingly important
role in science and technology. A number of different theoretical
techniques and approaches have been recently proposed on the path
to a unified theory\cite{"Michel:"2003",Bennaceur:2000,EPPEL:1975}.
In this presentation we discuss
simple examples that stress the complexity of the structure-reaction
borderline physics. We concentrate on the effects that the intrinsic structure
of composite objects plays in quantum-mechanical scattering and tunneling
processes. These processes are governed by non-perturbative physics
with exponential sensitivity to various quantities. We examine
two examples which, in our view, illuminate the physics of interest
from two opposite sides. The first example is a model of the one-dimensional scattering of 
a two-body system from a potential, and the second case
is a realistic example of $^{6}$He-neutron scattering. Despite complexity
and non-perturbative nature we obtain an exact quantum-mechanical
solution to the model example, while in the realistic case the problem
is solved with a Continuum Shell Model approach\cite{Volya:2005PRL,volya:2007}
using an effective Hamiltonian. Both our examples highlight similar
features related to the interplay of internal structure and reaction channels
and emphasize the roles of symmetries and unitarity.

\section{Reactions with a composite object: a model study}

As our model example we consider a one-dimensional scattering problem
where a projectile is a composite object made up of two particles
bound by a potential $v(x_{1}-x_{2})$ which depends only on the relative
distance between the particles. Here the particle coordinates are
$x_{1}$ and $x_{2}$, and the masses are $m_{1}$ and
$m_{2}$, respectively. This composite system interacts with an external
scattering potential $V$. We assume that
the external potential acts only on the second particle so that it
depends only on the coordinate $x_{2}.$ Below we use the usual center-of-mass
coordinate $R$ and relative coordinate $r=x_{1}-x_{2}$ and assume $M$
and $\mu$ to be the total and reduced masses of the system, respectively.
In these coordinates the total Hamiltonian is \[
H=-\frac{1}{2M}\frac{\partial^{2}}{\partial R^{2}}+h+V\,,
\quad{\rm where}\quad h=-\frac{1}{2\mu}\frac{\partial^{2}}{\partial r^{2}}+v(r)\]
 is the intrinsic Hamiltonian of the system.

The channels $|n\rangle$ are the eigenstates of the intrinsic Hamiltonian: 
$h|n\rangle=\epsilon_{n}|n\rangle$, so that the asymptotic forms for the 
incoming-plus-reflected and transmitted waves are \begin{equation}
|\Psi_{-}\rangle=e^{iK_{0}R}|0\rangle+\sum_{n=0}^{\infty}C_{-,n}e^{-iK_{n}R}|n\rangle,\, 
\text{and}\,\, \quad|\Psi_{+}\rangle=\sum_{n=0}^{\infty}C_{+,n}e^{iK_{n}R}|n\rangle.\label{refl wave}\end{equation}
Here $|\Psi_{-}\rangle$ includes the incoming wave in $n=0$ channel, 
the ground state of the intrinsic potential; $K_{n}$ is the channel momentum
$K_{n}(E)=\sqrt{2M(E-\epsilon_{n})}$ at a given total energy $E.$
The sums in (\ref{refl wave})
implicitly contain both open and closed channels depending on whether
the corresponding $K_{n}(E)$ is real or purely imaginary. In the
expression we use the principal value of the square root so that the reflected
waves in closed channels exponentially fall off. The conservation
of the center-of-mass flux in the open channels leads to the unitarity
relation
\begin{equation}
\sum_{n\,{\rm open}}\left(R_{n}+T_{n}\right)=1,\,\,\text{where}\,\,\, 
R_{n}=\frac{K_{n}}{K_{0}}\,|C_{-,n}|^{2}\,\,\,\,\text{and}\,\,\,\, 
T_{n}=\frac{K_{n}}{K_{0}}\,|C_{+,n}|^{2}\label{eq:flux}\end{equation}
are, respectively, the reflection and transmission probabilities in the $n$-th
channel. 

We model the external potential with a delta-peak, and only the second
particle is assumed to interact with it: $V(x_{2})=\alpha\delta(x_{2}).$
We implement a usual treatment of a delta-potential by separating
left and right regions denoted by $-$ and $+$ subscripts, respectively.
The principal complication in this problem comes from the boundary
condition on $x_{2}$ being incompatible with the center-of-mass coordinates.
The boundary conditions projected onto the $m$-th quantum state result
in a system of linear equations: \begin{equation}
\sum_{n}[C_{+,n}\langle m|D(k_{n})|n\rangle-C_{-,n}\langle m|D(-k_{n})|n\rangle]
=\langle m|D(k_{0})|0\rangle,\label{eq:D1}\end{equation}
$$
\sum_{n}\left [C_{+,n}\langle m|Q(k_{n})-2m_{2}\alpha D(k_{n})|n\rangle-C_{-,n}\langle m|Q(-k_{n})|n\rangle \right]=
\langle m|Q(k_{0})|0\rangle.$$
 Here for simplicity of notations we use an intrinsic momentum shift
operator $D(k)=e^{ikr}$ and the operator $Q(k)=i[\rho kD(k)-D(k)p]$
where $p=-i{\partial}/{\partial r}\,$ and $\rho={m_{2}}/{m_{1}}.$

We show our results for a case where the binding potential is given
by that of a harmonic oscillator. The expectation value of the momentum
shift operator can be expressed analytically with the Associated Laguerre
Polynomials. We choose $\omega$ as our
energy scale leaving relative kinetic energy $\mathcal{E}=E/\omega-1/2$ and relative
energy scale of delta-peak $\Delta={M\alpha^{2}}/{\omega}$ as
energy parameters. 

The problem expressed by equations \eqref{eq:D1} 
is that of an infinite set of linear equations which had to be solved by truncating
the space and including only a finite number of intrinsic states.
The momentum-shifts for highly virtual channels occur along the imaginary axis; 
therefore the momentum-shift operator matrix elements are exponentially divergent for these channels,
 and so are the corresponding coefficients $C_{\pm,n}$.  This shows a mathematically complex behavior near the barrier
where the boundary conditions are satisfied by cancellations of exponentially
divergent terms. Physically, however, highly-virtual excitations decay
fast leading to a regular behavior away from the delta peak.
Thus, this well-formulated problem of quantum mechanics appears to
be quite challenging mathematically. The proof of validity of the truncation 
mentioned above is an important issue addressed in Ref.\cite{"naureen:2007"}. 
The flux conservation \eqref{eq:flux} provides
an additional test of consistency and convergence. 

\begin{figure}[ht]
\vskip -0.6 cm
%\epsfxsize=10cm   %width of figure - will enlarge/reduce the figures
%\epsfbox{fig3.eps}
%\figurebox{2cm}{3cm}{} %to have a box alone
\begin{minipage}[1]{3.0 in}
\centerline{\epsfxsize=3.0in\epsfbox{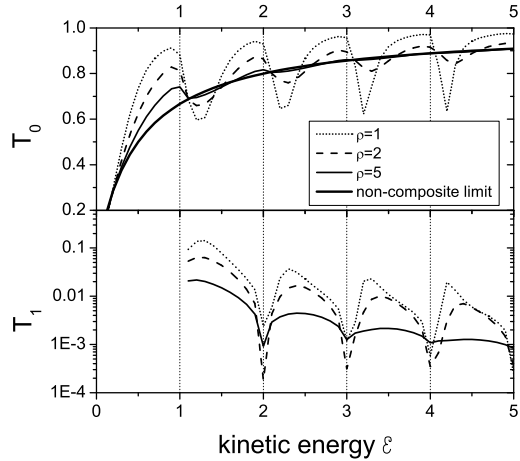}}
\end{minipage}
\begin{minipage}{1.2 in}
\caption{Transmission probabilities for the first two channels of a composite particle through the delta
barrier as a function of kinetic energy $\mathcal{E}$ at $\Delta=1$
with different $\rho$'s.\label{fig:Transmission}}
\end{minipage}
\vskip -0.6 cm
\end{figure}

The transmission probabilities for the two lowest channels resulting
from the scattering of an oscillator-bound system off a delta function
are shown in Fig.\ref{fig:Transmission}. The figure demonstrates
some of the generic features inherent to the composite-particle scattering.
The incident wave contains a composite particle in the ground state
and the number of open reflection/transmission channels depends
on the incident kinetic energy. For harmonic oscillator the threshold
energies for channel-opening correspond to integral values of $\mathcal{E}$.
At low energies, $\mathcal{E}<1$, transmission and reflection only
in the ground state are possible, and $T_{0}+R_{0}=1.$ Once the $\mathcal{E}=1$
threshold is crossed transmission and reflection in the first excited
oscillator state are also possible and the total flux is then shared
among all four processes: $T_{0}+T_{1}+R_{0}+R_{1}=1$. The number
of open channels increases with each integral value of $\mathcal{E}.$
The redistribution of probabilities at the threshold values leads
to cusps in the cross sections\cite{Landau:1981,Baz:1971}. A careful examination
of Fig.\ref{fig:Transmission} reveals the appearance of such sharp points
at thresholds. In addition to these, an interesting resonant-type
behavior can be noted in the transmission (and reflection, not shown
here) probabilities associated with peaks that do not coincide with
threshold energies. This resonant behavior is related to the intrinsic
structure. The case of a non-composite particle is a standard textbook
example, where the reflection $R=(1+2[\mathcal{E}/\Delta])^{-1}$ 
depends only on kinetic energy relative to the delta strength. In the figure, 
this non-composite limit for the corresponding kinematic conditions is 
shown with a thick solid line that has no oscillations. Within
our model this limit can be continuously reached when $\rho\rightarrow\infty$,
namely, when the mass of a non-interacting particle approaches zero. %

For the mass ratio $\rho$ of the orders of $1$ or smaller the internal structure
 plays an increasingly important role in the dynamics. The
sensitivity of $R$ and $T$ to the parameters of the model becomes
large and there appear resonances associated with the internal structure.

\section{Interplay of structure and reactions in the unified continuum shell
model approach}

We study our second, realistic example of $^{7}$He+n scattering 
using a Continuum Shell Model approach which is discussed in a series
of recent publications\cite{Volya:2003PRC,Volya:2005PRL}. The projection
formalism that lies in the foundation of the model dates back to 
Feshbach\cite{Feshbach:1958}.  The detailed study can be found
in the textbooks\cite{Mahaux:1969,Feshbach:1991}, and the previous
development of the method was reviewed by Rotter\cite{Rotter:1991}.
This method is actively used in diverse areas of open quantum many-body
systems from molecular and condensed matter physics\cite{Volya:2003JOB}
to multi-quark systems\cite{Auerbach:2004}. We review some
of the important ingredients below.

\subsection{Structure\label{structure}}

The part of the Hilbert space related to the particle(s) in the continuum
is eliminated with the projection formalism. As a result the ``intrinsic''
dynamics is given by the effective Hamiltonian \begin{equation}
{\cal H}(E)=H_{0}+\Delta(E)-\frac{i}{2}W(E).\label{1}\end{equation}
 Here the full Hamiltonian $H_{0}$ is restricted to the intrinsic space,
and is supplemented with the Hermitian term $\Delta(E)$ that describes
virtual particle excitations into the excluded space. The imaginary term
$i W(E)/2$ represents irreversible decays to the continuum. These new terms
in the projected Hamiltonian (\ref{1}) are given in terms of the matrix elements
$A_{1}^{c}(E)=\langle1|H_{0}|c;E\rangle$ of the full original Hamiltonian 
that link the internal states $|1\rangle$
with the energy-labeled external states $|c;E\rangle$ in the following manner:
\begin{equation}
\Delta_{12}(E)={\rm P.v.}\,\int dE'\sum_{c}\frac{A_{1}^{c}(E')A_{2}^{c\ast}(E')}{E-E'},\,\, W_{12}(E)
=2\pi\sum_{c({\rm open})}A_{1}^{c}A_{2}^{c\ast}.\label{3}\end{equation}

The properties of the effective Hamiltonian (\ref{1}) are as follows:

{\it 1.}  For unbound states, above the decay thresholds, the effective Hamiltonian
is non-Hermitian which reflects the loss of probability from the intrinsic
space.
 
{\it 2.}  The Hamiltonian has explicit energy dependence, making the internal
dynamics highly non-linear.

{\it 3.}  The solution for each individual nucleus is coupled to all the daughter systems via a chain of reaction channels. 

{\it 4.}  Even with two-body forces in the full space, the many-body interactions
appear in the projected effective Hamiltonian.

{\it 5.}  The eigenvalue problem ${\cal H}(E)|\alpha\rangle={\cal E}|\alpha\rangle$ 
represents a condition for the many-body resonant Siegert states,
for which the regular wave function is matched with the purely outgoing one at 
infinity. Below all decay thresholds the 
imaginary part disappears and the problem is equivalent to that of a traditional
shell model. Above decay thresholds, in general, there are no real energy
solutions, i.e., the stationary state boundary condition cannot be
satisfied. The complex energy eigenvalues correspond to poles of the
scattering matrix.

\subsection{Reactions and Unitarity}

The picture where the nuclear system is probed from {}``outside\char`\"{}
is given by the transition matrix defined within the general scattering
theory\cite{Mahaux:1969}, \begin{equation}
T^{ab}(E)=\sum_{12}A_{1}^{a\ast}(E)\,\left(\frac{1}{E-{\cal H}(E)}\right)_{12}A_{2}^{b}(E).\label{7}\end{equation}
 The poles of this transition matrix and the related full scattering matrix
$S=1-2\pi iT$ are the eigenvalues of the effective Hamiltonian. The
reaction theory is fully consistent with the structure description in Sec.~\ref{structure}.
However, many-body complexity, numerous poles, overlapping resonances
and energy dependence can make the observable cross-section quite
different from a collection of individual resonance peaks.

The transition matrix (\ref{7}) with the dimensionality equal to
the number of open channels can be written as $T={\bf A}^{\dagger}{\cal G}{\bf A}$,
where the full effective Green's function ${\cal G}(E)={1}/({E-{\cal H}})$
includes the loss of probability into all decay channels. 
The non-Hermitian part of Eq.(\ref{1}) is factorized as
$W=2\pi{\bf A}{\bf A}^{\dagger},$ where ${\bf A}$ represents a 
channel matrix (a set of column-vectors $A_{1}^{c}$ for all the channels $c$).
As shown in Refs.\cite{Durand:1976,Sokolov:1989}, 
iteration of the Dyson equation using the definitions ${\cal H}=H-iW/2$
and $G=(E-H)^{-1}$ leads to the following transition and scattering
matrices \begin{equation}
T=\frac{R}{1+i\pi R}\,,\quad S=\frac{1-i\pi R}{1+i\pi R}.\label{9}\end{equation}
 The matrix $R={\bf A}^{\dagger}G{\bf A}$ is analogous to the
$R$-matrix of the standard reaction theory; it is based on the Hermitian
part of the Hamiltonian $H=H_{0}+\Delta$. Thus, the factorized nature of
the intrinsic Hamiltonian and the appearance of the same effective
operator in the scattering matrix are important consequences of
unitarity.

\subsection{The He example}

\begin{figure}[ht]
\vskip -0.1 cm
%\epsfxsize=10cm   %width of figure - will enlarge/reduce the figures
%\epsfbox{fig3.eps}
%\figurebox{2cm}{3cm}{} %to have a box alone
\begin{minipage}[1]{2.6 in}
\centerline{\epsfxsize=2.8in\epsfbox{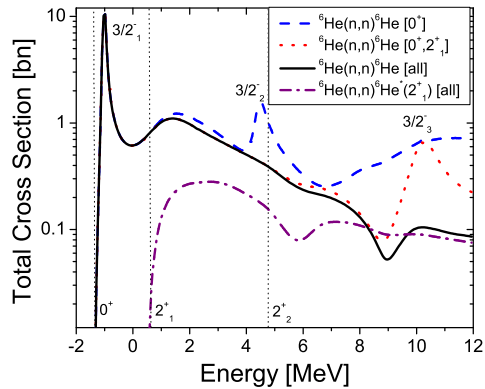}}
\end{minipage}
\begin{minipage}{1.8 in}
\caption{Cross-section for the neutron scattering of $^{6}$He in $0^{+}$ ground state. 
The solid curve is the total elastic cross-section, 
while the dashed and dotted curves correspond to the cases when
only $0^{+}$ channel (final state of $^{6}$He) and both $0^{+}$ and 
$2_{1}^{+}$ channels are included,
respectively. The dash-dot curve shows inelastic cross-section with
$^{6}$He in the $2_{1}^{+}$ final state. Thresholds for
the lowest three channels, $0^{+}$, $2_{1}^{+}$ and $2_{2}^{+}$, are
marked with vertical grid lines. Locations for three $3/2^{-}$ resonances
in $^{7}$He are indicated.\label{fig:he7}}
\end{minipage}
\vskip -0.4 cm
\end{figure}

In Fig.\ref{fig:he7} we consider a realistic example of $^{6}$He-neutron 
scattering. The parameters of the model are given by the
intrinsic shell model Hamiltonian from Ref. \cite{volya:2007} within the p-shell valence space. 
The continuum reaction physics
is modelled by the Woods-Saxon Hamiltonian. The model is discussed
in-depth in Ref.\cite{volya:2007}, where the entire chain of He
isotopes is solved in a coupled manner. For simplification of this discussion
we consider states in $^{6}$He to be bound, and concentrate on the role
of the internal structure in scattering with a neutron. Energies are
quoted here relative to the alpha particle ground state.

{\it 1.}  At low energies the peak in the cross-section corresponds to a narrow
resonance, $E_{r}$=-1.02 MeV, and a width of 91 keV which corresponds well
to a spectroscopic factor $C^2S$= 0.498.

{\it 2.}  The threshold to a second decay-channel (decay to a $2^{+}, 1.89 MeV$
excited state in $^{6}$He) is at 0.515 MeV. At this energy there is
a cusp in the cross-section; however, for a p-wave neutron the curve is smooth unlike that for an s-wave.

{\it 3.}  The resonance corresponding to the second $3/2^{-}$ state (5.510 MeV
excitation energy) in $^{7}$He appears at 4.494 MeV only 
when all other decay channels are ignored (dashed line in Fig.\ref{fig:he7}).
This state has a large width to decay into
$2^{+}$ final state in $^{6}$He, the $C^2S=$1.03 which makes resonance
peak impossible to observe.

\section{Conclusions}

In this work we target the issue of internal degrees of freedom in
scattering processes. We use two models, which are very different
in their nature and nicely show different aspects of the physics of
interest. The exactly solved simple one-dimensional scattering shows
an unusual resonant behavior associated with the composite nature
of the incident system. An application of the Continuum Shell Model
was demonstrated and discussed within a realistic example. Interplay
between channels, unitarity, and distribution of flux are common to
both examples and lead to generic near-threshold behavior. We emphasize
the importance of future theoretical developments toward a unified description
of structure and reactions.

%\bibliographystyle{h-elsevier}
%\bibliography{visnp}

\end{document}